\documentclass[twoside,twocolumn,aps,prb,superscriptaddress,longbibliography]{revtex4-2}
\usepackage[latin9]{inputenc}
\usepackage{amsmath}
\usepackage{amssymb}
\usepackage{graphicx}
\PassOptionsToPackage{version=3}{mhchem}
\usepackage{mhchem}

\makeatletter

\usepackage{float}
\usepackage{soul}
\usepackage{changes}
\usepackage{mathrsfs}
\usepackage{comment}
\usepackage{multirow}
\usepackage{xcolor}

\definecolor{kc}{rgb}{0.6,0,0.6}

\makeatother

\begin{document}
\title{Spin relaxation in graphite due to spin-orbital-phonon interaction from first-principles density-matrix approach}
\setcounter{page}{1}
\date{\today}
\author{Junqing Xu}
\email{jqxu@hfut.edu.cn}
\affiliation{Department of Physics, Hefei University of Technology, Hefei, Anhui, China}
\begin{abstract}
We predict ``intrinsic'' spin relaxation times ($T_{1}$) of graphite
due to spin-orbit-phonon interaction, i.e., the combination of spin-orbit
coupling and electron-phonon interaction, using our developed first-principles
density-matrix approach. We obtain ultralong $T_{1}$, e.g., $\sim$600
ns at 300 K, which leads to ultralong in-plane spin diffusion length
$\sim$110 $\mu$m within the drift-diffusion model. Our prediction
sets the upper bound of $T_{1}$ of graphite at each given temperature
and Fermi level. The anisotropy ratios of $T_{1}$ or values of $T_{1z}/T_{1x}$
are found small and around 0.6. We examine the applicability of the
well-known Elliot-Yafet (EY) relation, which declares that spin relaxation
rate $T_{1\alpha}^{-1}$ ($\alpha=x,y,z$) is proportional to the
product of the ensemble average of spin mixing parameter $\left\langle b_{\alpha}^{2}\right\rangle $
and carrier relaxation rate $\tau_{p}^{-1}$. Our numerical tests
suggest that the EY relation works qualitatively if the degeneracy
threshold $t^{\mathrm{deg}}$ for evaluating $b_{\alpha}^{2}$ is
relatively large (not much smaller than or comparable to $k_{B}T$),
e.g., $10^{-3}$ eV or larger, but fails if $t^{\mathrm{deg}}$ is
too tiny (much smaller than $k_{B}T$), e.g., $10^{-6}$ eV or smaller.
\end{abstract}
\maketitle

\section{Introduction}

Spin lifetime $\tau_{s}$, including spin relaxation time $T_{1}$
and (ensemble) spin dephasing time ($T_{2}^{*}$) $T_{2}$, and spin
diffusion length $l_{s}$ are key parameters for spintronic device
applications, which aim to achieve the next generation of low-power
electronics by making use of the spin degree of freedom.\citep{vzutic2004spintronics,wu2010spin}
In spintronic devices, $\tau_{s}$ and $l_{s}$ of materials in spin
transport channels are typically required long enough for stable detection
and manipulation of spin. Due to the weak spin--orbit coupling (SOC)
and negligible hyperfine interaction, carbon materials and nanomaterials
such as graphene\citep{avsar2020colloquium}, carbon nanotubes\citep{hueso2007transformation}
and graphite nanostructures\citep{banerjee2010spin}, are expected
to have long $\tau_{s}$ and $l_{s}$ and have been considered as
promising candidates of spintronic materials. Up to $\sim$10 $\mu$s
$\tau_{s}$, corresponding to hundreds of $\mu$m $l_{s}$, was predicted
for perfect suspended graphene.\citep{habib2022electric,xu2020spin}
However, much shorter experimental values of $\tau_{s}$, 100 ps to
12 ns at room temperature, were reported, due to extrinsic origins
such as impurities, substrates, etc.\citep{avsar2020colloquium,habib2022electric}

Recently, ultralong 100 ns $T_{1}$ of graphite at room temperature
was obtained in ESR measurements by Markus et al. in Ref. \citenum{markus2023ultralong}.
Since $2T_{1}$ sets up the upper bound of $\tau_{s}$ and $l_{s}$
is known proportional to $\tau_{s}$,\citep{vzutic2004spintronics,wu2010spin}
the knowledge of $T_{1}$ is critical to spintronic applications.
Remarkably, they also found highly anisotropic $T_{1}$ with $T_{1z}/T_{1x}$
up to 10, which is advantageous to efficient control of spin transport.
However, due to the existence of various non-magnetic and magnetic
impurities in realistic samples and possible Larmor precession induced
by the orbital Zeeman effect in ESR measurements, the experimental
data of $T_{1}$ may be considerably lower than the ideal or intrinsic
values due to spin-orbit-phonon interaction, i.e., the combination
of spin-orbital coupling and electron-phonon (e-ph) interaction. Therefore,
theoretical predictions of intrinsic $T_{1}$ are urgently needed
to guide the future improvements of $T_{1}$.

Moreover, in the work of Markus et al., the simple Elliot-Yafet (EY)
relation\citep{fabian1998spin,vzutic2004spintronics} was employed
to analyze the anisotropy of $T_{1}$. The EY relation declares that
$T_{1\alpha}^{-1}$ ($\alpha=x,y,z$) is proportional to the product
of the ensemble average of spin mixing parameter $\left\langle b_{\alpha}^{2}\right\rangle $
and carrier relaxation rate $\tau_{p}^{-1}$, and typically $T_{1\alpha}^{-1}\sim4\left\langle b_{\alpha}^{2}\right\rangle \tau_{p}^{-1}$.
Hereinafter, we name such quanlitative estimate of $T_{1}$ as $T_{1\alpha}^{\mathrm{EY}}$.
Based on the EY relation, the authors concluded that the high anisotropy
of $T_{1}$ is possibly attributed to the high anisotropy of spin-mixing
parameter $b^{2}$. However, they only evaluated $b_{kn}^{2}$ (state-resolved
$b^{2}$) at some k-points along the $M$-$K$-$\Gamma$ path, where
the anisotropy of $b_{kn}^{2}$ varies from 0.5 to \textasciitilde$3\times10^{6}$,
but had not obtained the Fermi-surface averaged values $\left\langle b^{2}\right\rangle $.
More crucially, the applicability of the EY relation to graphite is
problematic due to the following issue: The band structure of graphite
contains multiple band-crossing points and four-fold degenerate regions
near Fermi surface. Since $b_{kn}^{2}$ is evaluated within the degenerate
subspace, the value of $b_{kn}^{2}$ is sensitive to the choice of
the degeneracy threshold $t^{\mathrm{deg}}$, if the state $\left(k,n\right)$
is close to the band-crossing point or in a near-degeneracy region.
Therefore, $\left\langle b_{\alpha}^{2}\right\rangle $ and accordingly
$T_{1\alpha}^{\mathrm{EY}}$ becomes strongly $t^{\mathrm{deg}}$
dependent, which is unphysical. Additionally, the EY relation is derived
based on the assumptions of small $b^{2}$ and slow variation of $b_{kn}^{2}$
over wavevector, which are however not satisfied in their work. Therefore,
it is urgent to simulate $T_{1}$ directly via advanced \textit{ab initio}
methods and examine the applicability of the EY relation.

\begin{figure*}
\includegraphics[scale=0.49]{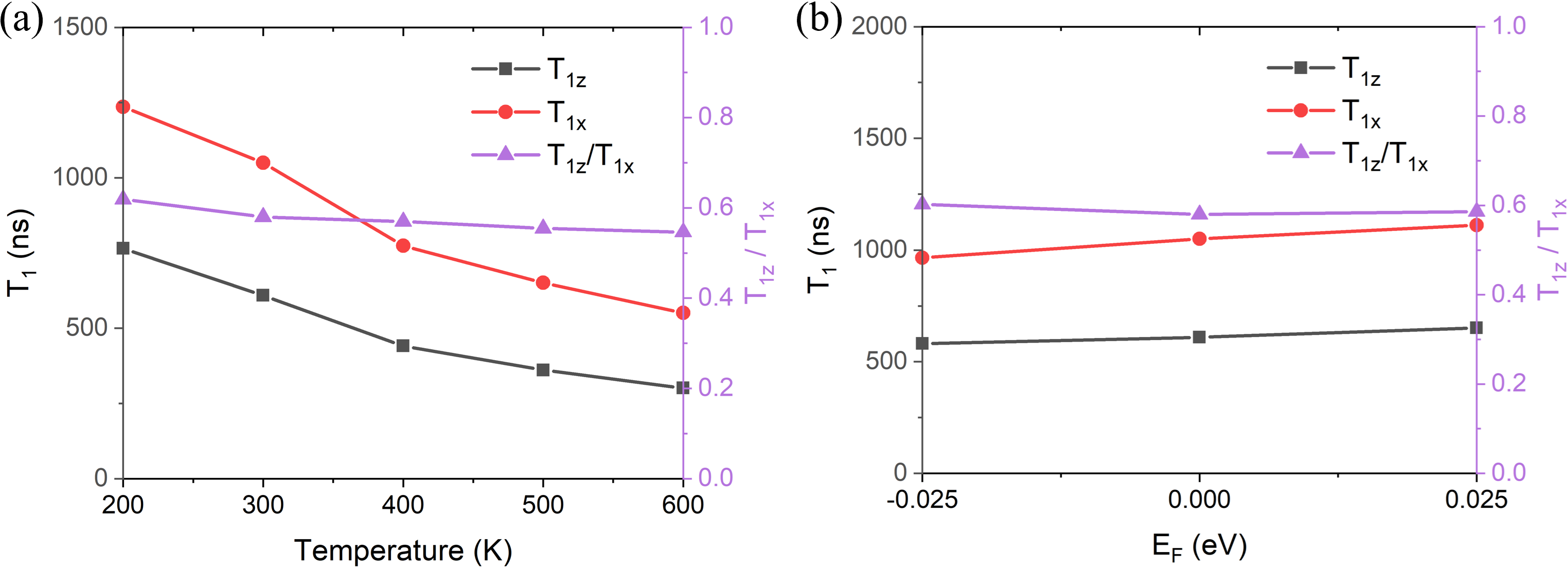}

\caption{Theoretical results of $T_{1}$ and its anisotropy of graphite calculated
by FPDM approach (a) at different temperatures with Fermi level $E_{\mathrm{F}}$=0
and (b) with different $E_{\mathrm{F}}$ at 300 K.\label{fig:T1}}
\end{figure*}

In this work, we apply our developed first-principles density-matrix
dynamics (FPDM) approach\citep{xu2023ab,xu2021ab,xu2024ab} with self-consistent
SOC and quantum description of the electron-phonon scattering, to
simulate intrinsic $T_{1}$ of graphite due to spin-orbit-phonon interaction
at various temperatures and Fermi levels ($E_{F}$). FPDM approach
was applied to disparate materials including silicon, (bcc) iron,
transition metal dichalcogenides (TMDs), graphene-hBN, GaAs and CsPbBr$_{3}$,
in good agreement with experiments.\citep{xu2023ab,xu2024spin} We
then compare our FPDM results of $T_{1}$ with results from the EY
relation - $T_{1}^{\mathrm{EY}}$, to gain mechanistic insights of
spin relaxation in graphite.

\section{Methods}

\subsection{The density matrix master equation and spin lifetime}

We solve the quantum master equation of density matrix $\rho\left(t\right)$
as the following:\citep{rosati2014derivation,xu2021ab}
\begin{align}
\frac{d\rho_{12}\left(t\right)}{dt}= & -\frac{i}{\hbar}\left[H_{e},\rho\left(t\right)\right]_{12}+\nonumber \\
 & \left(\begin{array}{c}
\frac{1}{2}\sum_{345}\left\{ \begin{array}{c}
\left[I-\rho\left(t\right)\right]_{13}P_{32,45}\rho_{45}\left(t\right)\\
-\left[I-\rho\left(t\right)\right]_{45}P_{45,13}^{*}\rho_{32}\left(t\right)
\end{array}\right\} \\
+H.C.
\end{array}\right),\label{eq:master}
\end{align}
Eq.~\ref{eq:master} is expressed in the Schr\"odinger picture,
where the first and second terms on the right side of the equation
relate to the coherent dynamics and the scattering processes within
Born-Markov approximation respectively. $H_{e}$ is the electronic
Hamiltonian. $\left[H,\rho\right]\,=\,H\rho-\rho H$. H.C. is Hermitian
conjugate. The subindex, e.g., ``1'' is the combined index of k-point
and band. The weights of k-points must be considered when doing sum
over k-points. $P$ is the generalized scattering-rate matrix for
the e-ph scattering, computed from the e-ph matrix elements and energies
of electrons and phonons. Details of $P$ are given in Appendix A.

Starting from an initial density matrix $\rho\left(0\right)$ prepared
with a net spin, we evolve $\rho\left(t\right)$ through Eq. \ref{eq:master}
for a long enough time. We then obtain the excess spin observable
$S^{\mathrm{ex}}\left(t\right)$ from $\rho\left(t\right)$ and extract
$T_{1}$ from $S^{\mathrm{ex}}\left(t\right)$. See details in Appendix
B.

\subsection{Spin-mixing parameter $b^{2}$ and the EY relation\label{subsec:b2_and_EY_relation}}

Due to time-reversal and spatial inversion symmetries of graphite
(and each layer of it), every two bands of graphite form a Kramers
degenerate pair,\citep{vzutic2004spintronics} so that spin-up/down
is well defined along an arbitrary axis ${\bf \widehat{r}}$ by diagonalizing
the corresponding spin matrix $s_{{\bf r}}={\bf s}\cdot{\bf \widehat{r}}$
in degenerate subspaces. Therefore, spin relaxation in graphite is
dominated by EY mechanism, i.e., caused by spin-flip transitions.
For EY spin relaxation, $T_{1}^{-1}$ is often well described by Fermi's-golden-rule-like
(FGR-like) formulas (Eqs. \ref{eq:FGR-comm} and \ref{eq:FGR} in
Appendix C.2) and qualitatively satisfies the EY relation described
below.

\subsubsection{Spin-mixing parameter $b^{2}$}

Suppose the spin of a state ``1'' is highly polarized along $i$
direction. Then in general, the wavefunction of state ``1'' can
be written as $\Psi_{1}\left({\bf r}\right)=a_{\alpha,1}\left({\bf r}\right)\alpha+b_{\alpha,1}\left({\bf r}\right)\beta$,
where $a$ and $b$ are the coefficients of the large and small components
of the wavefunction, and $\alpha$ and $\beta$ are spinors (one up
and one down for direction $i$). Define $a_{i,1}^{2}=\int|a_{i,1}\left({\bf r}\right)|^{2}d{\bf r}$
and $b_{\alpha,1}^{2}=\int|b_{\alpha,1}\left({\bf r}\right)|^{2}d{\bf r}$,
then $a_{\alpha,1}^{2}>b_{\alpha,1}^{2}$ and $b_{\alpha,1}^{2}$
is just spin-mixing parameter of state ``1'' along direction $\alpha$.\citep{xu2021giant,xu2023substrate}
\begin{align}
a_{\alpha,1}^{2}+b_{\alpha,1}^{2}= & 1,\\
0.5\left(a_{\alpha,1}^{2}-b_{\alpha,1}^{2}\right)= & \left|S_{\alpha,1}^{\mathrm{exp}}\right|,
\end{align}

where $S_{\alpha,1}^{\mathrm{exp}}$ is the spin expectation value
of state ``1'' along the direction $\alpha$, and is an eigenvalue
of the degeneracy projection of $s_{\alpha}$ within the degenerate
subspace containing state ``1''. From the above equations,
\begin{align}
b_{\alpha,1}^{2}= & 0.5-\left|S_{\alpha,1}^{\mathrm{exp}}\right|.
\end{align}

Since $S^{\mathrm{exp}}$ depends on the degeneracy threshold $t^{\mathrm{deg}}$,
$b^{2}$ is also $t^{\mathrm{deg}}$ dependent. According to our discussions
in Appendix C.3, a too small $t^{\mathrm{deg}}$ (such as 10$^{-6}$
eV or smaller) may be problematic in certain cases, while a larger
$t^{\mathrm{deg}}$ (such as 10$^{-3}$ eV or larger) leads to reasonable
results.

\subsubsection{The EY relation}

For EY spin relaxation, $T_{1}^{-1}$ may be well described by FGR
with spin-flip transitions. Thus, $T_{1}^{-1}$ is qualitatively proportional
to $\left|g^{\uparrow\downarrow}\right|^{2}$ with $g^{\uparrow\downarrow}$
the spin-flip scattering matrix element. Similarly, $\tau_{p}^{-1}\propto\left|g^{\uparrow\uparrow}\right|^{2}$
with $g^{\uparrow\uparrow}$ the spin-conserving scattering matrix
element. Therefore, we have
\begin{align}
T_{1}^{-1}/\tau_{p}^{-1}\propto & \left|g^{\uparrow\downarrow}\right|^{2}/\left|g^{\uparrow\uparrow}\right|^{2}.
\end{align}

According to Refs. \citenum{fabian1998spin,leyland2007oscillatory,restrepo2012full},
$\left|g^{\uparrow\downarrow}\right|^{2}/\left|g^{\uparrow\uparrow}\right|^{2}\sim b^{2}$,
so that
\begin{align}
T_{1}^{-1}/\tau_{p}^{-1}\sim & \left\langle b^{2}\right\rangle .\label{eq:taus_b2_relation}
\end{align}
Practically, we use the following approximate relation\citep{fabian1998spin}
\begin{align}
T_{1}^{-1}/\tau_{p}^{-1}= & 4\left\langle b^{2}\right\rangle .\label{eq:EY_relation}
\end{align}

This is called the EY relation in this work. Note that the EY relation
is a rough relation and may lead to huge errors for some materials.\citep{xu2023substrate,xu2024spin}

\section{Results and discussions}

\subsection{$T_{1}$ from FPDM approach}

\begin{figure*}
\includegraphics[scale=0.72]{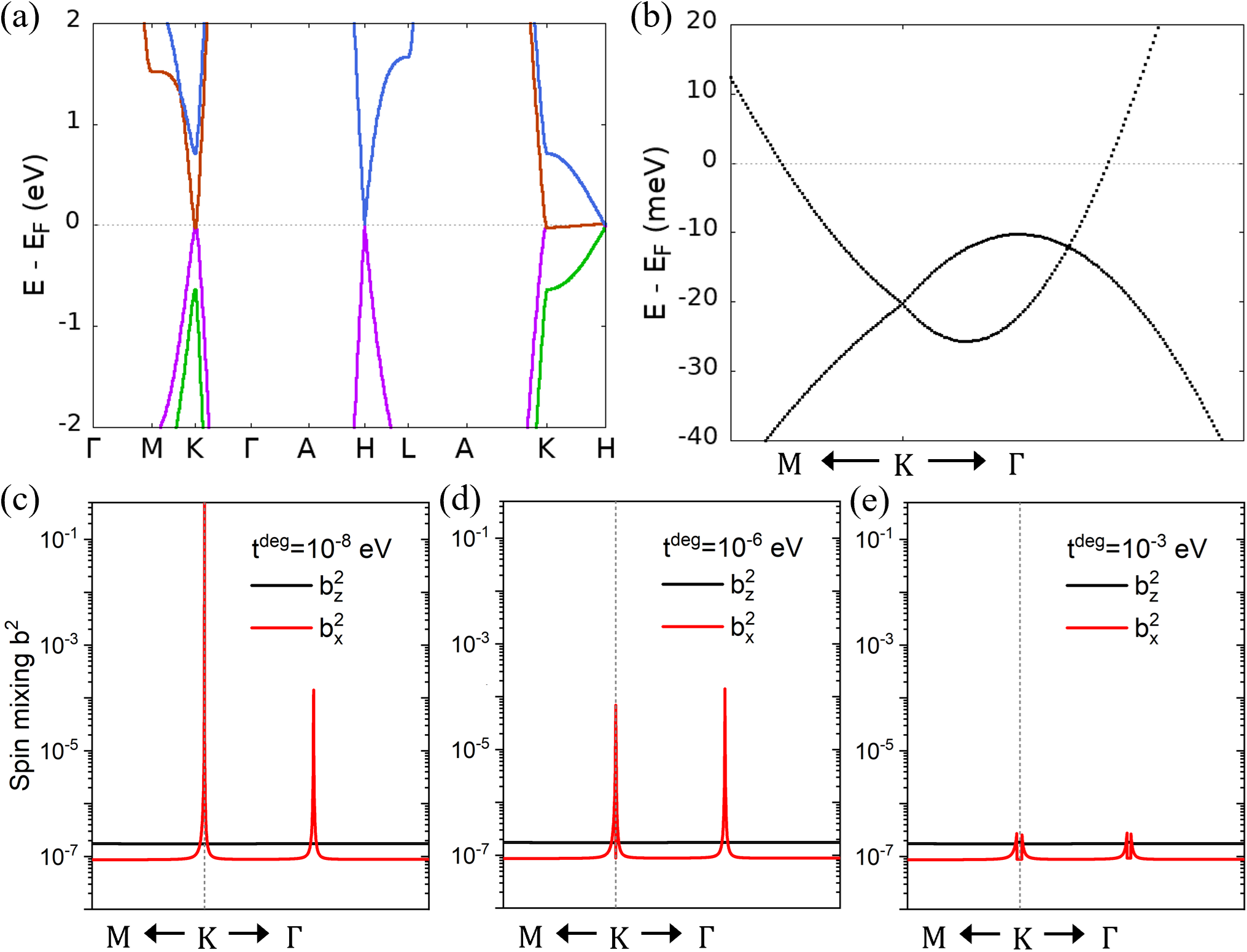}

\caption{(a) Band structure of graphite. (b) Band structure around $K$ along
$M$-$K$-$\Gamma$. (c) The k-resolved spin mixing parameter $b_{\alpha,k}^{2}$
calculated with the degeneracy threshold $t^{\mathrm{deg}}$=10$^{-8}$
eV of k-points around $K$ along $M$-$K$-$\Gamma$. $b_{k}^{2}$
is defined as $b_{\alpha,k}^{2}=\frac{\sum_{n}f'_{kn}b_{\alpha,kn}^{2}}{\sum_{n}f'_{kn}}$,
where $f'_{kn}$ is the derivative of the Fermi-Dirac distribution
function of state $\left(k,n\right)$ and $b_{\alpha,kn}^{2}$ is
the state-resolved spin-mixing parameter. (d) and (e) are $b_{\alpha,k}^{2}$
calculated with $t^{\mathrm{deg}}$ of 10$^{-6}$ and 10$^{-3}$ eV
respectively.\label{fig:b2k}}
\end{figure*}

We first show $T_{1}$ results calculated by FPDM approach at different
temperatures and $E_{F}$ in Fig. \ref{fig:T1}. We obtain ultralong
$T_{1}$ in a wide range of temperatures and $E_{F}$. At 300 K and
$E_{F}$=0, $T_{1z}$ and $T_{1x}$ can reach 609 and 1050 ns respectively,
if we only consider spin relaxation due to spin-orbit-phonon interaction.
The magnitude of $T_{1}$ is found decreasing with temperature and
increasing slightly with $E_{F}$. On the other hand, the anisotropy
ratio of $T_{1}$ - $T_{1z}/T_{1x}$ is found around 0.55-0.6 and
insensitive to temperature and $E_{F}$. More analysis of theoretical
results of $T_{1}$ are given below in the next subsection.

In recent ESR experiments\citep{markus2023ultralong}, $T_{1z}\sim100$
ns and $T_{1x}\sim10$ ns at room temperature were reported. Thus,
our FPDM approach predicts much longer $T_{1}$ but much smaller anisotropy
ratio ($T_{1z}/T_{1x}$). The differences may be attributed to the
following reasons: (i) There exist non-magnetic and magnetic impurities
in experimental samples. The existence of the impurities will reduce
the magnitude of $T_{1}$ and possibly change the anisotropy of $T_{1}$.
(ii) Finite magnetic fields are applied in ESR measurements. Due to
the orbital Zeeman effect, corresponding to the Hamiltonian $H_{k}^{\mathrm{Z,orb}}=\mu_{B}{\bf B}\cdot{\bf L}_{k}$
with ${\bf L}_{k}$ the orbital angular momentum operator matrix,
k-dependent Larmor precession processes may be induced by external
magnetic field.\citep{xu2024spin} These processes possibly cause
an additional spin relaxation channel. The effects of impurities and
$H_{k}^{\mathrm{Z,orb}}$ on spin lifetimes are however hard to simulate
and beyond the scope of this work.

Within the drift-diffusion model, spin-diffusion length of $s_{\alpha}$
along the direction $\beta$ of a non-magnetic metal can be estimated
as $l_{s_{\alpha}\beta}=\sqrt{\tau_{p}v_{F\beta}^{2}T_{1\alpha}}$,\citep{markus2023ultralong}
where $v_{F\beta}$ is the Fermi velocity along the direction $\beta$.
Thus, with \textit{ab initio} results of $\tau_{p}$ (196 fs) and
$v_{F\beta}$ ($v_{Fx}\approx3.4\times10^{5}$ m/s), we obtain ultralong
in-plane spin diffusion lengths $l_{s_{z}x}\sim$117 $\mu$m and $l_{s_{x}x}\sim$154
$\mu$m at 300 K. These values are several times longer than the reported
experimental $l_{s}$ of graphene up to $\sim$30 $\mu$m at room
temperature\citep{drogeler2016spin,avsar2020colloquium} and make
graphite a promising candidate of spintronic materials. The out-of-plane
spin diffusion lengths are shorter and our estimates are $l_{s_{z}z}\sim$12
$\mu$m and $l_{s_{x}z}\sim$16 $\mu$m at 300 K with theoretical
$v_{Fz}\approx3.5\times10^{4}$ m/s.

\subsection{Results related to the EY relation}

\begin{figure*}
\includegraphics[scale=0.69]{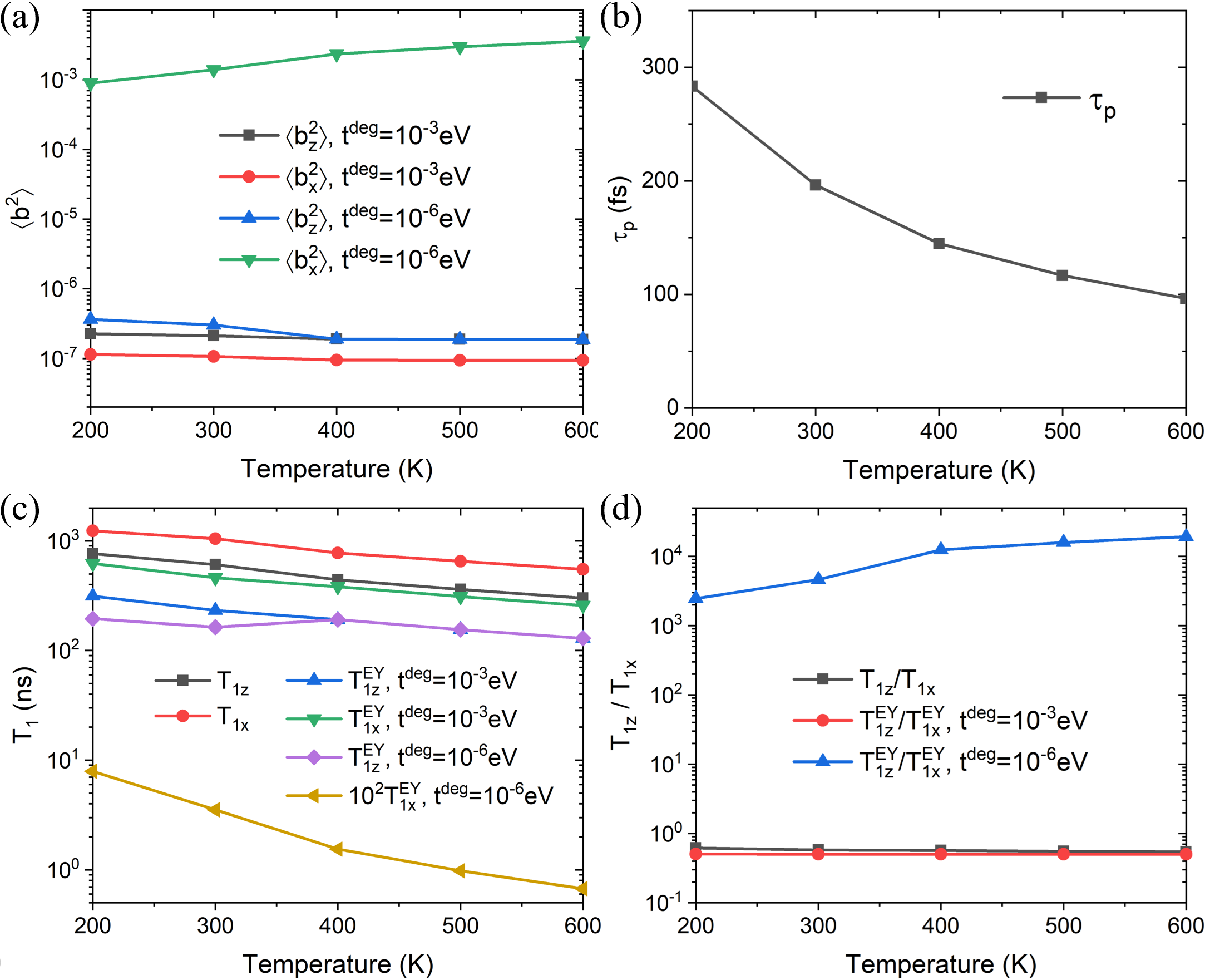}

\caption{Theoretical analysis related to the EY relation. (a) The Fermi-surface
averaged spin mixing parameter along the directions $z$ and $x$
- $\left\langle b_{z}^{2}\right\rangle $ and $\left\langle b_{x}^{2}\right\rangle $
as a function of temperature with different $t^{\mathrm{deg}}$. (b)
Carrier lifetime $\tau_{p}$. (c) Spin relaxation times from FPDM
approach - $T_{1z}$ and $T_{1x}$, compared with those estimated
by the EY relation (Eq. \ref{eq:EY_relation}) - $T_{1z}^{\mathrm{EY}}$
and $T_{1x}^{\mathrm{EY}}$ with different $t^{\mathrm{deg}}$. Note
that $T_{1x}^{\mathrm{EY}}$ values with $t^{\mathrm{deg}}$=10$^{-6}$
eV are multiplied by 100. (d) The anisotropy ratios of $T_{1}$ and
$T_{1}^{\mathrm{EY}}$ with different $t^{\mathrm{deg}}$.\label{fig:EY}}
\end{figure*}

As discussed in Sec. \ref{subsec:b2_and_EY_relation}, spin relaxation
in graphite at zero magnetic field should be caused by EY mechanism
and can be qualitatively described by the EY relation (Eq. \ref{eq:EY_relation}).
Therefore, to improve the understandings of FPDM results of $T_{1}$
(Fig. \ref{fig:T1} and the above subsection), we next study the spin
mixing parameter $b^{2}$ and $T_{1}$ results estimated by the EY
relation.

In Fig. \ref{fig:b2k}, we show band structure and the calculated
k-resolved spin mixing parameter $b_{\alpha,k}^{2}$ ($\alpha=x,z$)
of graphite. $b_{\alpha,k}^{2}$ is the band average of $b_{\alpha,kn}^{2}$
at a k-point (See the caption of Fig. \ref{fig:b2k}). Due to time-reversal
and spatial inversion symmetries, every two bands of graphite form
a Kramers degenerate pair.\citep{vzutic2004spintronics} The band
structure of graphite is found rather complicated: There are 8 bands
near $E_{F}$. Moreover, there exist multiple band-crossing points,
which are four-fold degenerate, near the $K$-$H$ line. For example,
along the $M$-$K$-$\Gamma$ line, there are two band-crossing points
- one very close to $K$ and another at $K_{2}$ a little away from
$K$.

According to the discussions in Sec. \ref{subsec:b2_and_EY_relation},
$b_{\alpha,kn}^{2}$ is computed from $s_{\alpha}^{\mathrm{deg}}$
- the degeneracy projection of the spin operator matrix $s_{\alpha}$
within the degenerate subspace containing the state $\left(k,n\right)$.
Therefore, $b_{\alpha,k}^{2}$ can be $t^{\mathrm{deg}}$ dependent.
Indeed, from Fig. \ref{fig:b2k}(c)-(e), we find that $b_{x,k}^{2}$
are very sensitive to $t^{\mathrm{deg}}$ near $K$ and $K_{2}$ (band-crossing
points shown in Fig. \ref{fig:b2k}(b)), although $b_{z,k}^{2}$ at
all k-points and $b_{x,k}^{2}$ at other k-points (away from $K$
and $K_{2}$) are stable against $t^{\mathrm{deg}}$ and close to
$8.8\times10^{-8}$ and 1.75$\times$10$^{-7}$ respectively. With
$t^{\mathrm{deg}}\le10^{-6}$ eV, the curve of $b_{x,k}^{2}$ has
two sharp peaks at $K$ and near $K_{2}$ with the peak amplitudes
$\ge10^{-4}$. With $t^{\mathrm{deg}}=10^{-8}$ eV, the maximum value
of $b_{x,k}^{2}$ is quite large and only slightly small than 0.5
(at $K$). Our results of $b_{\alpha,k}^{2}$ with $t^{\mathrm{deg}}=10^{-8}$
eV are almost the same as Ref. \citenum{markus2023ultralong}. The
large magnitudes of $b_{x,k}^{2}$ indicate that $s_{x,k}$ cannot
be safely replaced by $s_{x,k}^{\mathrm{deg}}$ with $t^{\mathrm{deg}}\le10^{-6}$
eV, since the off-diagonal element of $s_{x,k}$ is large between
two states with a finite but tiny energy difference ($\le10^{-6}$
eV). On the other hand, with relatively large $t^{\mathrm{deg}}$
of 10$^{-3}$ eV, $b_{z,k}^{2}$ and $b_{x,k}^{2}$ at all k-points
are tiny ($<$3$\times$10$^{-7}$), indicating that $s_{x,k}\approx s_{x,k}^{\mathrm{deg}}$
if $t^{\mathrm{deg}}\ge10^{-3}$ eV.

In Fig. \ref{fig:EY}(a), we show Fermi-surface averaged values $\left\langle b_{\alpha}^{2}\right\rangle $
with different $t^{\mathrm{deg}}$. It can be seen that with relative
large $t^{\mathrm{deg}}$ of 10$^{-3}$ eV, $\left\langle b_{z}^{2}\right\rangle $
and $\left\langle b_{x}^{2}\right\rangle $ are of order $10^{-7}$
and not very sensitive to temperature (and also $E_{F}$ as checked).
However, with $t^{\mathrm{deg}}=10^{-6}$ eV, $\left\langle b_{x}^{2}\right\rangle $
become much larger and of order $10^{-3}$, while $\left\langle b_{z}^{2}\right\rangle $
are still small. See the additional details concerning the k-point
convergence of $\left\langle b_{\alpha}^{2}\right\rangle $ simulations
in Appendix D and Fig. \ref{fig:b2_convergence}.

With $\left\langle b_{\alpha}^{2}\right\rangle $ and \textit{ab initio}
$\tau_{p}$ shown in Fig. \ref{fig:EY}(b) due to the e-ph scattering,
we estimate spin relaxation times using the EY relation, called $T_{1}^{\mathrm{EY}}$
(Fig. \ref{fig:EY}(c)). With large $t^{\mathrm{deg}}$ of 10$^{-3}$
eV, we find that the EY relation qualitatively captures the magnitude
and temperature dependence of $T_{1}$, i.e., $T_{1}^{\mathrm{EY}}$
and $T_{1}$ from FPDM approach have the same order of magnitude at
different temperatures and their temperature dependences are quite
similar. Since $\left\langle b_{\alpha}^{2}\right\rangle $ is weakly
temperature dependent, the temperature dependences of $T_{1}$ and
$\tau_{p}$ can be attributed to the same reason: With increasing
temperature, the phonon occupations are increased, so that both the
spin-flip and spin-conserving e-ph scattering strengths are enhanced,
which correspond to larger $T_{1}^{-1}$ and $\tau_{p}^{-1}$. On
the other, with tiny $t^{\mathrm{deg}}$ of 10$^{-6}$ eV, $T_{1x}^{\mathrm{EY}}$
values are found faster than FPDM values of $T_{1x}$ by 3-4 orders
of magnitude and the temperature dependence of $T_{1z}^{\mathrm{EY}}$
is not consistent with FPDM $T_{1z}$. Moreover, from Fig. \ref{fig:EY}(d),
we find that $T_{1z}^{\mathrm{EY}}/T_{1z}^{\mathrm{EY}}$ are in agreement
with and completely different from FPDM $T_{1z}/T_{1x}$ when $t^{\mathrm{deg}}$
is $10^{-3}$ and $10^{-6}$ eV respectively. We have also checked
the EY relation with other $t^{\mathrm{deg}}$. As expected, we find
that $t^{\mathrm{deg}}$=10$^{-2}$ eV leads to quite similar results
to $t^{\mathrm{deg}}$=10$^{-3}$ eV, but $t^{\mathrm{deg}}$=10$^{-8}$
eV leads to unreasonable $T_{1}^{\mathrm{EY}}$.

Therefore, from Fig. \ref{fig:EY}(c) and (d), we conclude that the
EY relation works qualitatively with relatively large $t^{\mathrm{deg}}$
but fails with tiny $t^{\mathrm{deg}}$ for the zero-filed spin relaxation
in graphite due to the spin-orbit-phonon interaction. This conclusion
can be explained within the density-matrix master equation approach:
The use of the EY relation requires that spin relaxation is well described
by FGR with spin-flip transitions. According to Appendix C, to obtain
a FGR-type formula (Eq. \ref{eq:FGR}) from the density-matrix master
equation, it is required that $s_{\alpha}$ can be safely replaced
by $s_{\alpha}^{\mathrm{deg}}$ for bands near $E_{F}$, which ensures
that spin-up and spin-down are well defined for all electronic states
(in the new basis diagonalizing $s_{\alpha}^{\mathrm{deg}}$). However,
in general, the off-diagonal element of $s_{\alpha}$ between two
states with a finite energy difference is not negligible, especially
when the energy difference is small. Therefore, replacing $s_{\alpha}$
by $s_{\alpha}^{\mathrm{deg}}$ with a tiny $t^{\mathrm{deg}}$ may
lead to significant errors. A simple solution of this problem is increasing
$t^{\mathrm{deg}}$ and it seems reasonable to set $t^{\mathrm{deg}}$
comparable to or a bit smaller than $k_{B}T$ (see Appendix C.3).

\section*{Acknowledgments}

This work is supported by National Natural Science Foundation of China
(Grant No. 12304214), Fundamental Research Funds for Central Universities
(Grant No. JZ2023HGPA0291). This research used resources of the HPC
Platform of Hefei University of Technology.

\section*{Appendices}

\subsection*{Appendix A: The generalized scattering-rate matrix $P$}

We use the following form of $P$, which is called the Lindbladian
form here and reads\citep{xu2021ab,xu2023ab}
\begin{align}
P_{1234}= & \frac{2\pi}{\hbar}\sum_{q\lambda\pm}G_{13}^{q\lambda\pm}G_{24}^{q\lambda\pm,*}n_{q\lambda}^{\pm},\label{eq:P}\\
G_{13}^{q\lambda\pm}= & g_{12}^{q\lambda\pm}\sqrt{\delta_{\sigma}^{G}\left(\epsilon_{1}-\epsilon_{2}\pm\omega_{q\lambda}\right)},\label{eq:Gq}
\end{align}
where $q$ and $\lambda$ are phonon wavevector and mode, $g^{q\lambda\pm}$
is the e-ph matrix element, resulting from the absorption ($-$) or
emission ($+$) of a phonon, computed with self-consistent SOC from
first principles,\citep{giustino2017electron} $n_{q\lambda}^{\pm}=n_{q\lambda}+0.5\pm0.5$
in terms of phonon Bose factors $n_{q\lambda}$, and $\delta_{\sigma}^{G}$
represents an energy conserving $\delta$-function broadened to a
Gaussian of width $\sigma$.

According to Ref.~\citenum{rosati2014derivation}, there exists another
pathway to arrive at the Markov limit of the scattering term, leading
to another form of $P$, called the conventional form here. The conventional
form contains a single Dirac delta function and reads

\begin{align}
P_{1234}= & \sum_{q\lambda\pm}a_{13}^{q\lambda\pm}b_{24}^{q\lambda\pm,*},\label{eq:Peph_conv}\\
a_{13}^{q\lambda\pm}= & \frac{1}{\hbar}g_{13}^{q\lambda\pm}\sqrt{n_{q\lambda}^{\pm}},\\
b_{13}^{q\lambda\pm}= & 2\pi g_{13}^{q\lambda\pm}\sqrt{n_{q\lambda}^{\pm}}\delta\left(\omega_{13}\pm\omega_{q\lambda}\right).
\end{align}

The conventional form of $P$ was widely employed in previous theoretical
simulations of spin dynamics based on the density-matrix master equation
approach.\citep{wu2010spin} It however does not preserve the positive-definite
character of the occupation numbers or the diagonal elements of $\rho$,
unlike the Lindbladian form,\citep{rosati2014derivation} although
we find that two forms of $P$ lead to rather similar $\tau_{s}$
(within 10\% for graphite and within 20\% for various other materials)
and the same carrier lifetime $\tau_{p}$. Therefore, we will stick
to the Lindbladian form of $P$.

The smearing parameter $\sigma$ in the Lindbladian form has a weak
physical meaning and roughly corresponds to the collision duration
or $\tau_{p}$. $\sigma$ is set comparable to $k_{B}T$, which is
a few times larger than $\tau_{p}$, to speed up the k-point convergence.
Our numerical tests show that the resulting $T_{1}$ values are not
sensitive to $\sigma$. For example, the resulting $T_{1}$ is only
changed by a few percent when $\sigma$ is reduced to its half.

\subsection*{Appendix B: Spin lifetime}

$\rho\left(t\right)$ can be separated as
\begin{align}
\rho_{kmn}\left(0\right)= & f_{km}\delta_{mn}+\delta\rho_{kmn}\left(0\right),
\end{align}

where $f_{km}$ is the Fermi-Dirac occupation of state $\left(k,m\right)$.
$\delta\rho\left(t\right)$ is the excess density matrix and typically
small in a spin-relaxation simulation.

Suppose the initial time is $t=0$, starting from an initial excess
density matrix $\delta\rho\left(0\right)$ prepared with a net excess
spin along the direction $\alpha$, we evolve $\rho\left(t\right)$
through Eq. \ref{eq:master} for a long enough time, typically a few
ns. We then obtain the excess spin observable along the direction
$\alpha$ - $S_{\alpha}^{\mathrm{ex}}\left(t\right)$ from $\delta\rho\left(t\right)$
through
\begin{align}
S_{\alpha}^{\mathrm{ex}}\left(t\right)= & \mathrm{Tr}\left[\delta\rho\left(t\right)s_{\alpha}\right].
\end{align}

Then $T_{1\alpha}$ is extracted by fitting $S_{\alpha}^{\mathrm{ex}}\left(t\right)$
using the exponential decay curve:
\begin{align}
S_{\alpha}^{\mathrm{ex}}\left(t\right)= & S_{\alpha}^{\mathrm{ex}}\left(0\right)\mathrm{exp}\left(-\frac{t}{T_{1}}\right).\label{eq:exp_decay}
\end{align}

As we have checked, the results of $T_{1\alpha}$ are almost unchanged
if the length of time evolution is increased from a few ns to a few
hundreds of ns. This is because the time evolution of $S_{\alpha}^{\mathrm{ex}}\left(t\right)$
fits Eq. \ref{eq:exp_decay} excellently even at the first few ps
for graphite.

\subsection*{Appendix C: The approximate formula of $T_{1}$ and FGR}

An alternative form of the initial excess density matrix $\delta\rho\left(0\right)$
is
\begin{align}
\delta\rho_{kmn}\left(0\right)= & \delta\rho_{kmn}^{s_{\alpha}},\label{eq:drho0}\\
\delta\rho_{kmn}^{s_{\alpha}}= & \frac{\mu_{B}g_{0}\delta B}{\hbar}s_{\alpha,kmn}\left(\frac{\Delta f}{\Delta\epsilon}\right)_{kmn},\label{eq:drho_dB}\\
\left(\frac{\Delta f}{\Delta\epsilon}\right)_{kmn}= & f'_{km}\delta_{\epsilon_{km}\epsilon_{kn}}+\frac{f_{km}-f_{kn}}{\epsilon_{km}-\epsilon_{kn}}\left(1-\delta_{\epsilon_{km}\epsilon_{kn}}\right),\label{eq:deltaf_by_deltae}
\end{align}

where $\delta\rho^{s_{\alpha}}$ is the density-matrix change induced
by a spin Zeeman Hamiltonian $H^{\mathrm{sZ}}=\mu_{B}g_{0}\delta Bs_{\alpha}$
with a perturbation magnetic field $\delta B$ along the direction
$\alpha$.\citep{xu2020spin} $s_{\alpha}$ is spin Pauli matrix in
Bl\"och basis along direction $\alpha$.

\subsubsection*{Appendix C.1: The approximate formula of $T_{1}$ with $\delta\rho^{s_{\alpha}}$}

Due to the inversion symmetry of graphite (and each layer of it),
there are no SOC-induced band splittings and no k-dependent effective
SOC fields (called ``internal magnetic fields''). Therefore, Larmor
spin precession, caused by the coherent term of the master equation
Eq. \ref{eq:master}, should be insignificant. Considering that spin
relaxation is mainly determined by the scattering strength and Larmor
precession frequency,\citep{vzutic2004spintronics,wu2010spin} the
coherent term is probably unimportant to spin relaxation in graphite
and can be approximately neglected.

Without Larmor precession, we may assume that the excess density matrix
$\delta\rho\left(t\right)$ is approximately proportional to $\delta\rho^{s_{\alpha}}$
given above (Eq. \ref{eq:drho_dB}) at any $t$, if initially we take
$\delta\rho\left(0\right)=\delta\rho^{s_{\alpha}}$. This is a nature
choice of $\delta\rho\left(t\right)$ for the simulation of $T_{1\alpha}$
due to several reasons: (i) The instantaneous spin relaxation rates
$T_{1}^{-1}$ at any $t$ are the same if $\delta\rho=\lambda\delta\rho^{s_{\alpha}}$
with $\lambda$ a constant and $\lambda\le1$. (ii) $\delta\rho^{s_{\alpha}}$
leads to high spin polarizations along the direction $\alpha$ but
no excess charges at every k-point. (iii) The matrix elements $\delta\rho_{kmn}$
are large and negligible when states $\left(k,m\right)$ and $\left(k,n\right)$
are both near and far from $E_{F}$ respectively, consistent with
the facts that $T_{1}$ is a Fermi-surface property and the e-ph scattering
processes are only significant with partial occupations. Therefore,
$T_{1\alpha}$ can be approximately computed from
\begin{align}
\frac{dS_{\alpha}^{\mathrm{ex}}}{dt}= & -\frac{S_{\alpha}^{\mathrm{ex}}}{T_{1\alpha}},\\
S_{\alpha}^{\mathrm{ex}}= & \mathrm{Tr}\left[\delta\rho^{s_{\alpha}}s_{\alpha}\right].
\end{align}

Using the above equations and Eq. \ref{eq:master} without the coherent
term, we obtain the approximate formula of $T_{1\alpha}^{-1}$,
\begin{multline}
T_{1\alpha}^{-1}=\frac{2\pi}{\hbar N_{k}^{2}S_{\alpha}^{\mathrm{ex}}}\mathrm{Tr}_{n}\mathrm{Re}\sum_{kk'\lambda}\left[s_{\alpha},G^{q\lambda-}\right]_{kk'}\\
\times\left[\begin{array}{c}
\left(\delta\rho^{s_{\alpha}}\right)_{k}G_{kk'}^{q\lambda-}\left(n_{q\lambda}+I-f_{k'}\right)\\
-\left(n_{q\lambda}+f_{k}\right)G_{kk'}^{q\lambda-}\begin{array}{c}
\left(\delta\rho^{s_{\alpha}}\right)_{k'}\end{array}
\end{array}\right]^{\dagger_{n}}.\label{eq:approximate_T1}
\end{multline}

Here, the $G$ is exactly as defined above in Eq.~\ref{eq:Gq}, but
separating the wave vector indices ($k$, $k'$) and writing it as
a matrix in the space of band indices ($n$,$n'$) alone. Similarly,
$s_{\alpha}$ and $\delta\rho^{s_{\alpha}}$ are also matrices in
the band space, $\mathrm{Tr}_{n}$ and $\dagger_{n}$ are trace and
Hermitian conjugate in band space, and $[o,G]_{kk'}\equiv o_{k}G_{kk'}-G_{kk'}o_{k'}$,
written using matrices in band space. Eq. \ref{eq:approximate_T1}
is the same as Eq. 3 in Ref. \citenum{xu2020spin}.

\subsubsection*{Appendix C.2: FGR}

If the bands near $E_{F}$ are all degenerate or the matrix elements
$s_{\alpha,kmn}$ with $\epsilon_{km}\neq\epsilon_{kn}$ are all negligible.,
the spin matrix $s_{\alpha}$ in $\delta\rho^{s_{\alpha}}$ (Eq. \ref{eq:drho_dB})
can be safely replaced by its degenerate-subspace projection $s_{\alpha}^{\mathrm{deg}}$,
whose matrix elements satisfy $s_{\alpha,kmn}^{\mathrm{deg}}=s_{\alpha,kmn}\delta_{\epsilon_{km}\epsilon_{kn}}$.
In such cases, we can further simplify Eq.~\ref{eq:approximate_T1}
to the FGR-like expression (the derivation is given in SI of Ref.
\citenum{xu2020spin}),
\begin{multline}
T_{1\alpha}^{-1}=\frac{2\pi}{\hbar N_{k}^{2}S_{\alpha}^{\mathrm{ex}}k_{B}T}\sum_{kk'\lambda\pm nn'}\left\{ \left|\left[s_{\alpha}^{\mathrm{deg}},g^{q\lambda-}\right]_{knk'n'}\right|^{2}\right.\\
\left.\delta\left(\epsilon_{kn}-\epsilon_{k'n'}-\omega_{q\lambda}\right)f_{k'n'}\left(1-f_{kn}\right)n_{q\lambda}\right\} .\label{eq:FGR-comm}
\end{multline}

In the new basis obtained by diagonalizing $s_{\alpha}$ or $s_{\alpha}^{\mathrm{deg}}$
within degenerate subspaces, the above equation becomes
\begin{multline}
T_{1\alpha}^{-1}=\frac{2\pi}{\hbar N_{k}^{2}S_{\alpha}^{\mathrm{ex}}k_{B}T}\times\\
\sum_{kk'\lambda\pm nn'}\left\{ \left|\left(S_{\alpha,kn}^{\mathrm{exp}}-S_{\alpha,k'n'}^{\mathrm{exp}}\right)g_{knk'n'}^{q\lambda-}\right|^{2}\right.\\
\left.\delta\left(\epsilon_{kn}-\epsilon_{k'n'}-\omega_{q\lambda}\right)f_{k'n'}\left(1-f_{kn}\right)n_{q\lambda}\right\} ,\label{eq:FGR}
\end{multline}

where $S_{\alpha,kn}^{\mathrm{exp}}$ is the spin expectation value
along the direction $\alpha$ and $S_{\alpha,kn}^{\mathrm{exp}}\equiv s_{\alpha,knkn}$
(in the new basis).

In many systems such as conduction bands of silicon and monolayer
MoS$_{2}$, the values of $S_{\alpha}^{\mathrm{exp}}$ are either
$\approx0.5\hbar$ or $\approx-0.5\hbar$. In such systems, according
to Eq. \ref{eq:FGR}, only the spin-flip transitions between two states
with opposite signs of $S_{\alpha}^{\mathrm{exp}}$ have non-negligible
contributions to $T_{1\alpha}^{-1}$. Then, Eq. \ref{eq:FGR} becomes
a FGR-type formula with initial and final states having opposite signs
of $S_{\alpha}^{\mathrm{exp}}$, corresponding to spin-flip transitions.

\begin{figure}[t]
\includegraphics[scale=0.38]{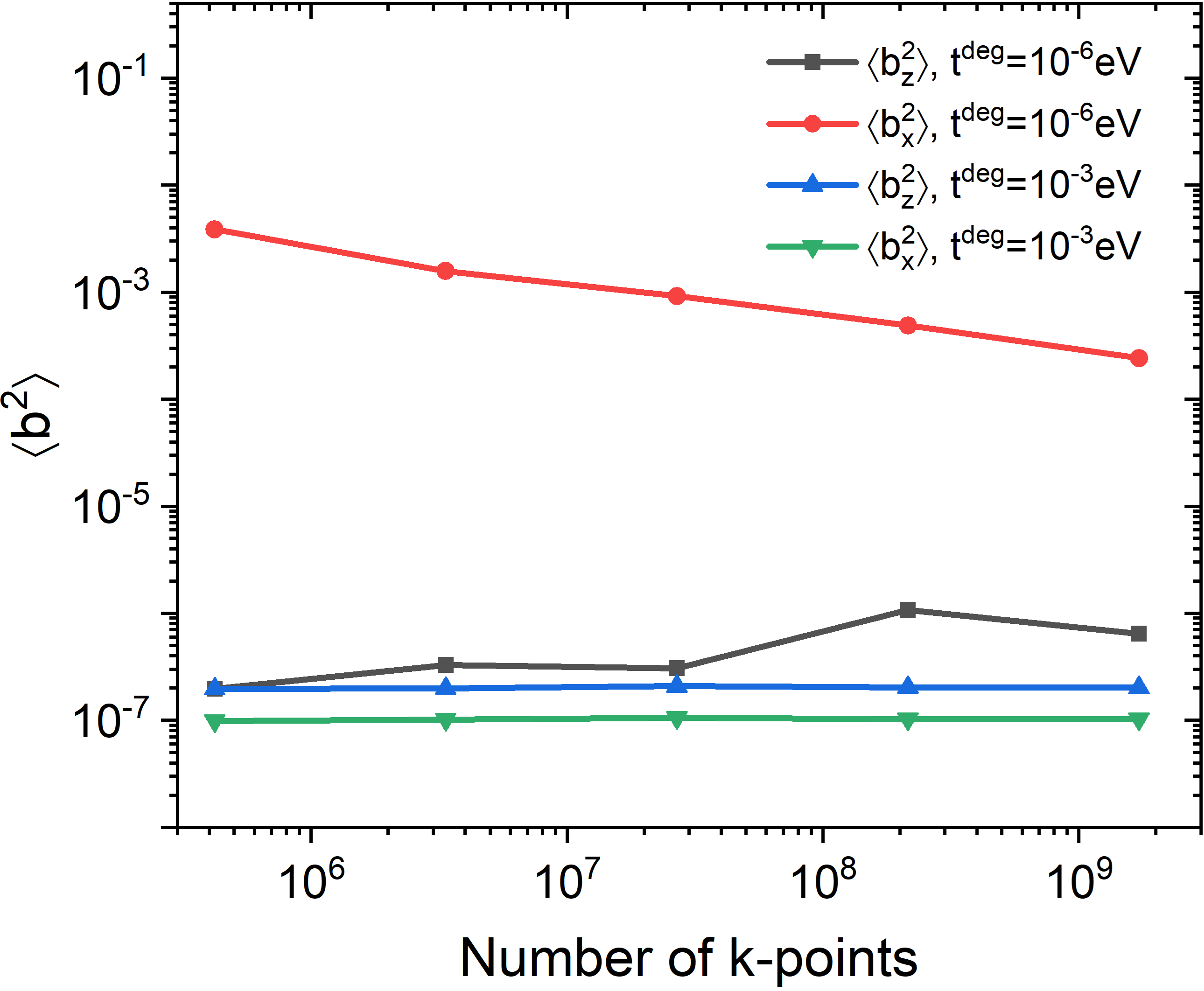}

\caption{$\left\langle b^{2}\right\rangle $ as a function of number of k-points.
The uniform k meshes $18N\times18N\times6N$ with $N$ an integer
are used for $\left\langle b^{2}\right\rangle $ simulations.\label{fig:b2_convergence}}
\end{figure}

\subsubsection*{Appendix C.3: The degeneracy threshold}

When the coherent dynamics is unimportant to spin relaxation, \textit{ab initio}
results of $T_{1}$ calculated by Eq. \ref{eq:approximate_T1} with
$\delta\rho^{s_{\alpha}}$ have been found close to the FPDM results
(Eq. \ref{eq:master}) for various materials including graphite. However,
the FGR-like formulas (Eqs. \ref{eq:FGR-comm} and \ref{eq:FGR})
may lead to significant errors, if the degeneracy threshold $t^{\mathrm{deg}}$
is too small. This is because replacing $s_{\alpha}$ in $\delta\rho^{s_{\alpha}}$
(Eq. \ref{eq:drho_dB}) by $s_{\alpha}^{\mathrm{deg}}$ with too small
$t^{\mathrm{deg}}$ is problematic in some cases, where the matrix
elements $s_{\alpha,kmn}$ with $\epsilon_{km}\neq\epsilon_{kn}$
are not negligible. In such cases, when $\epsilon_{km}$ and $\epsilon_{kn}$
are both near $E_{F}$ and $\left|\epsilon_{km}-\epsilon_{kn}\right|$
is a few $k_{B}T$ or smaller, the matrix element $\left(\frac{\Delta f}{\Delta\epsilon}\right)_{kmn}$
in $\delta\rho^{s_{\alpha}}$ is not much smaller than $\frac{df}{d\epsilon}|_{\epsilon=\overline{\epsilon}}$
with $\overline{\epsilon}=\frac{\epsilon_{km}+\epsilon_{kn}}{2}$,
so that neglecting the off-diagonal matrix element $\delta\rho_{kmn}^{s_{\alpha}}$
is problematic and replacing $s_{\alpha}$ in $\delta\rho^{s_{\alpha}}$
by $s_{\alpha}^{\mathrm{deg}}$ with too small $t^{\mathrm{deg}}$
is improper.

Considering that $\left(\frac{\Delta f}{\Delta\epsilon}\right)_{kmn}$
varies relatively slow with the variation of $\left|\epsilon_{km}-\epsilon_{kn}\right|$
(in the scale of a few $k_{B}T$), a simple way to fix the above degeneracy
issue is using a relatively large $t^{\mathrm{deg}}$, comparable
to or a bit smaller than $k_{B}T$, for Eqs. \ref{eq:FGR-comm} and
\ref{eq:FGR}.

\subsection*{Appendix D: Computational details}

The ground-state electronic structure, phonons, as well as the e-ph
matrix elements are firstly calculated using density functional theory
(DFT) with relatively coarse $k$ and $q$ meshes in the DFT plane-wave
code JDFTx\citep{sundararaman2017jdftx}. We use the experimental
lattice constants with $a$=2.76 \AA and $c$=6.71 \AA,
which are close to the relaxed lattice constants. The exchange-correlation
functional is PBE\citep{perdew1996generalized}. The internal geometries
are fully relaxed using the DFT+D3 method for van der Waals dispersion
corrections\citep{grimme2010consistent}. We use Optimized Norm-Conserving
Vanderbilt (ONCV) pseudopotentials\citep{hamann2013optimized} with
self-consistent SOC throughout, which we find converged at a kinetic
energy cutoff of 74 Ry. The DFT calculations use $18\times18\times6$
$k$ meshes. The phonon calculation employs a $6\times6\times2$ supercell
through finite difference calculations.

We then transform all quantities from plane wave basis to maximally
localized Wannier function basis\citep{marzari1997maximally}, and
interpolate\citep{giustino2017electron} them to substantially finer
k and q meshes. The fine $k$ and $q$ meshes are typically $252\times252\times84$.
We have checked the k-point convergence carefully. $252\times252\times84$
k meshes are found fine enough to converge $\left\langle b^{2}\right\rangle $
results with $t^{\mathrm{deg}}\ge10^{-4}$ eV and $T_{1}$ results
within 5\% and 20\% respectively. Finer k meshes do not change the
trends of these results. However, we find that $\left\langle b^{2}\right\rangle $
results with $t^{\mathrm{deg}}\le10^{-6}$ eV are very sensitive to
k meshes and even $1728\times1728\times576$ k meshes are not enough
to converge them. We show $\left\langle b^{2}\right\rangle $ as a
function of number of k-points in Fig. \ref{fig:b2_convergence}.
It seems impractical to fully converge $\left\langle b^{2}\right\rangle $
results with $t^{\mathrm{deg}}\le10^{-6}$ eV. Our k-point convergence
tests indicate that tiny values of $t^{\mathrm{deg}}$ are not suitable
for numerical calculations of $\left\langle b^{2}\right\rangle $.
The real-time dynamics simulations are done with our own developed
DMD code interfaced to JDFTx.

\section*{Reference}


%

\end{document}